# Towards a theory of superplasticity

Egorushkin V.[a], Ponomarev A.[a,b]

[a] Institute of Strength Physics and Materials Science of SB RAS, 2/4 Akademichesky Av., Tomsk 634055, Russia

[b] National Research Tomsk Polytechnic University, 30 Lenin Avenue, 634050 Tomsk, Russia

E-mail: val110@mail.ru

**Abstract**

This work deals with an investigation of general principles of superplasticity (SP) in deformed materials. It is shown that a non-linear, wave plastic deformation is the basic process for all plastic deformation phenomena; it makes an individual contribution into these phenomena, allows describing them from the same standpoints, and offers a way to follow the relationship between the physics of defects and the mechanics of plasticity. It is to be noted that macro and meso defects – discontinuities in the vector fields of macro and meso elastic displacements, are no less fundamental than are the microdefects – dislocations and disclinations; it is the latter which form the process of localized macrodeformation. General mechanisms of this process are analyzed in this study.

Relying on the concepts of non-equilibrium thermodynamics, a constitutive equation of superplastic state is obtained, which relates the strain rate, its rotational modes, local irreversible stresses, temperature, and density of heat and mass transfer. Special cases and their derived relations are analyzed.

It is shown that SP is determined by plastic equilibrium concurrent with the composition and structure fluctuations.

An expression is obtained for the superplastic flow velocity, containing three terms: velocity of wave plastic macroprocess, increased by fluctuations, velocity prescribed by the gradient of external sources (diffusion, etc.), and velocity of the delayed intragranular slip. It should be noted that transformation is a characteristic and necessary feature of SP flow. In the absence of transformations, deformation occurs via creep.

A creep curve is calculated and the physical nature of its three-stage character is explained.

In addition, fragmentation of the deforming region and grain-boundary sliding (GBS) are discussed and so are their mechanisms and characteristic features. Formulas are derived for the strain rate under conditions of GBS and the reasons for its non-monotonic dependence on grain size, applied stresses and temperature are analyzed.

## 1. Introduction

A number of phenomena accompany plastic deformation of materials: grain boundary sliding, Coble and Nabarro–Herring creep, serrated yielding, and superplasticity, which exhibit common properties associated with grain-boundary deformation.

Plastic flow in all these phenomena develops under thermodynamic conditions along with inhomogeneous variations in composition or structure. Under non-equilibrium conditions (increased thermodynamic potential), inhomogeneities can develop only during the fluctuation-assisted surmounting of the energy barriers and system's motion towards equilibrium. Thus, it is along with the processes pushing the material to equilibrium that plastic flow occurs [1-3].

The effect of superplasticity – anomalous elongation and increased strain rate, in the initial stages is determined by grain-boundary sliding, while in the case of longer times and higher slip velocities – by grain-boundary and intragrain deformation [4]. Superplasticity is similar to creep, but unlike the latter it is characterized by smoothness of plastic flow, strong sensitivity to testing temperatures, and grain shape and size. It was reported that diffusion flows along the interfaces enhance the SP effect, while segregation of the diffusing components does not appreciably affect it [5].

It is accepted that SP and GBS rely on the dislocation mechanism of deformation: nucleation, motion, and absorption of grain-boundary and lattice dislocations. Note that the major contribution to SP comes from GBS, while diffusion-induced creep and intragranular slip play an accommodating role. Thus, the cause of SP rests in the behavior of microscopic defects.

Surprisingly, no one has ever tried to interpret the elastic mechanical properties of materials and structures from the perspective of the behavior of the field elasticity quanta. Similar to the classical displacement macrofields that play the leading role in elasticity mechanics, the macro (meso) discontinuities of these fields determine plastic macrodeformation.

Under the SP condition, the localized plastic deformation evolves along with the concentration and/or structure fluctuations, in cooperation with grain-boundary slip and intragranular deformation. As a result, the deforming material is in the conditions of elastoplastic (rather than thermodynamic) equilibrium until fracture.

To gain an insight into the above phenomena and to describe them properly, we have to address the dynamics of plastic deformation together with concentration fluctuations: to derive constitutive and equilibrium equations, as well as localized solutions of the latter.

Unlike the elastic equilibrium prescribed by the wave equation for elastic displacements and by the Hooke law – equation of state, plastic equilibrium is determined by the equation for discontinuities of elastic fields of displacements – their density and flow [6,7]. The equation of plastic state – a constitutive equation, is prescribed by a positive production of entropy controlled by the defect and energy flows from the local deforming regions. It is these flows which ensure irreversibility of the local deformation process.

Plastic displacements are certain 'compensating fields' for elastic deformations in a continuous medium and plastic flow dynamics is described by the calibrating fields or by the geometry of curved space-time.

The idea of using calibrating fields or Riemann geometry to describe defects in a deforming solid dates back to the 1980s [8]. Despite its long history, the discussion of microscopic equations well-known in the field theory, their symmetry and comparison of the field characteristics with the defects in order to clarify the 'physical meaning' did not advance much father [6,7]. The attempts to look for the reasons of a <u>macroscopic phenomenon</u> – plastic deformation, in microscopic mechanisms of atomic motion are successful in 'model specimens' only, having low density of defects – dislocations. Under real physical conditions, the contribution from a certain number of dislocations is unlikely to penetrate through the internal material structure and its effect can hardly be measured. The escape of a defect from the region of localization under straining controls the locomotive force and just ensures irreversibility of the process due to the entropy production rather than macroscopic effect of shape variation.

When we turn to structured media and meso- and macro-regions, we have to average microequations. This is necessarily done in mechanics and electrodynamics of continuous media.

Averaging of equations in the physics of defects (dislocations) was reported long ago [9], but similarly to the manner used in electrodynamics and with the same effect – development of dislocation-induced polarization.

From our perspective, it is the 'electrodynamic' result of averaging, which serves one of the reasons for describing macrodeformation via microdefects, taking into account their polarization. In electrodynamics, the medium and the magnetic field are different essences, and the problem of averaging is formulated with respect to the field. In physics (mechanics) of plasticity, the media and defects represent a single object. An averaging, therefore, implies variations in the scale of this object.

Scale averaging is well known in the mechanics of structured media. One of the procedures is based on scale transformations and ε-extension [10]. Using the data reported in [10], in [6,7] we averaged the microequations obtained and derived their macroanalog and localized extensions representing non-linear plastic waves of shape variation [6,7, 11].

This wave process is the basis for further consideration of the above phenomena; it makes an independent contribution into them, allows describing them from a single standpoint and makes it possible to follow the relationship of the physics of defects with the mechanics of plastic deformation.

## 2. Waves of plastic deformation

From the system of equations obtained in [6,7], which describe an elasto-plastic medium, let us take the equilibrium equation representing the dynamics of deformation. At the macroscale, it relates the variations in the flow and density of defects – discontinuities of the displacement vector field, $U_\mu$, and the elastic stresses and potential external sources of plastic deformation

$$\varepsilon_{\mu x \delta} \partial_x \alpha_\delta^\beta = \frac{1}{c^2} \partial_0 J_\mu^\beta + \sigma_\mu^\beta - \frac{c_{\mu\nu}^{\beta\gamma}}{E} p_\gamma^\nu, \qquad (1)$$

where $\varepsilon_{\mu x \delta}$ is the Levi-Civita symbol, $\partial_x$ is the derivative over the space variables, $\partial_0$ is the derivative with respect to time with summation performed over the reiterating indices, $\alpha_\delta^\beta$ is the linear defect density tensor (if the value of a discontinuity is a multiple of the crystal lattice constant, these defects are dislocations), $J_\mu^\beta$ is the plastic flow – defect flow density, $\sigma_\mu^\beta = \frac{c_{\mu\nu}^{\beta\gamma}}{E} \partial_\gamma \ln u_\nu$ are the local elastic stress concentrators determined by the Hooke law, and $p_\gamma^\nu$ is the plastic distortion having dual nature. On the one hand, $p_\gamma^\nu$ along with α and J determines the current deformation, on the other hand, together with $\sigma_\mu^\beta$ it can represent external sources (mechanisms) stimulating this deformation. Quantities $c_{\mu\nu}^{\beta\gamma}$ are the elastic moduli, E is the Young modulus, and c is the velocity of sound.

In the general case in (1) it seems impossible to get rid of $p_\gamma^\nu$, the calibration potential gradient, using calibration conditions similar to the electrodynamic case, since calibration is determined by the conditions of a specific problem.

In the case where elastic sources are determined by variation in the scalar $\psi(x,t)$ rather than displacement vector, which corresponds to the change in the volume of the deformed material fraction, the equation acquires a vector form

$$\varepsilon_{\mu x \delta} \partial_x \alpha_\delta = \frac{1}{c^2} J_\mu + \sigma_\mu - c_{\mu\nu}^\gamma p_\gamma^\nu, \qquad (2)$$

where $\sigma_\mu = c_\mu^\gamma \partial_\gamma \ln \psi(x,t)$, $c_\mu^\gamma = \sum_{\beta,\nu} \frac{c_{\mu\nu}^{\beta\gamma}}{E}$.

Equation (1) can also be given by the vector relation under a certain loading mode.

In order to transfer to the macroscopic relations, we have to perform scale averaging (1), (2). This averaging was carried out in [6,7] using scale transformations and epsilon-extension. The equation of macro (meso) plastic equilibrium for the above case is given by the following [6,7]:

$$V_\mu = \varepsilon_{\mu x \delta} \partial_x \Phi_\delta - c_{\mu\nu}^\gamma P_\gamma^\nu, \qquad (3)$$

where $\Phi = -c^2 \int \alpha \, dt$, $V = -J$, $P_\gamma^\nu = c^2 \int P_\gamma^\nu dt$, $c_{\mu\nu}^\gamma = \sum_\beta \frac{\langle c_{\mu\nu}^{\beta\gamma} \rangle}{E}$; the brackets indicate the space average [6,7] and time averaging is performed over the observation period. The formulas for deriving $\langle c_{\mu\nu}^{\beta\gamma} \rangle$ are given in [6,7]. The opposite signs of quantities $\Phi$ and V indicate that these are 'mass' parameters rather than defect characteristics.

The averaged equations possess a number of valuable properties:
1) Plastic deformation at the meso and macrolevels occurs without variations in the volume (similar to the case of microlevel deformation [9]) – following the volume conservation law. At the meso and macrolevels, volume variation vanishes during scale averaging.
2) In the general case of the vector source, the elastic shear stress concentrator does not vanish upon averaging, and the form of macroequations is similar to that of microequations (1). The averaged physical characteristics of the medium only – moduli of elasticity, density, etc., are subjected to variations. Thus, the dynamics of plastic deformation is similar for all of the scale (structural) levels, since it is expressed by similar equilibrium equations – following the scale similarity law.
3) The instantaneous velocity of plastic flow is determined by the characteristics of material imperfection, summarized over the period of deformation. These conclusions are implied by the most non-linear evolution equilibrium equation itself and are common for its any localized solutions. In fact, plastic shape variation is determined by the evolution of the local displacement field discontinuities. Therefore, it makes no sense to find solutions for (2), (3) in the entire space of coordinates. We have to find solutions to the wave equations in an isolated area undergoing deformation. It is clear that they cannot present waves in a general sense, but they have to describe shape variation of a local spatial region, whose axis represents a certain spatial curve. The theory of spatial curves is quite detailed; it is based on the Frenet equation, a local system of coordinates, and a possibility of reducing the equations of motion to non-linear equations solved by the method of an inverse scattering problem. This procedure was successfully implemented in [6,7].

The equations obtained for localized plastic deformation, (2), (3), demonstrated that deformation propagates in the form of <u>non-linear helical waves with the wave characteristics determined by the geometry (torsion curvature), scale (length, transverse dimensions), and physical (elastic modulus, density) parameters of the deforming regions [6,7]</u>.

It is reasonable to discuss these solutions in a greater detail for a number of reasons:
- Plastic deformation waves represent an independent fundamental phenomenon at all of the scale (structural) levels of deformation – macro, meso, and micro;
- The wave nature of plasticity underlies all plastic deformation phenomena;
- These solutions make an independent contribution into superplasticity and other above-mentioned types of deformation;
- They provide a relation between the physics of defects and mechanics of plastic deformation;
- They offer a possibility of analyzing the differences in the plastic flow of materials at different scale levels, from micro- to nanolevel.

Following the method proposed in [6,7], let us present the term $c_{\mu\nu}^{\gamma} p_{\gamma}^{\nu}$ as a sum of the rotor of a certain vector $\mathbf{\Phi}$ and the scalar gradient $f$

$$c_{\mu\nu}^{\gamma} p_{\gamma}^{\nu} = \varepsilon_{\mu x\delta}\partial_x \Phi'_{\delta} + \partial_{\mu} f, \qquad (4)$$

and take a rotor from both parts of equation (3). Then (3) will become

$$\varepsilon_{nk\mu}\varepsilon_{\mu x\delta}\partial_k \partial_x \Phi_{\delta} = W_n, \qquad (5)$$

where $W_n = \varepsilon_{nk\mu}\partial_k V_{\mu}$ is the vorticity of plastic flow and $\mathbf{\Phi}=\mathbf{\Phi}+\mathbf{\Phi}`$ contains the rotational part of the external plastic distortion. A solution of (5) in a local system of coordinates (normal $\mathbf{n}$, binormal $\mathbf{b}$, tangent $\mathbf{t}$) together with the Frenet equation, determines the velocity of plastic deformation, proportional to the curvature $x(S,t)$ of the deforming region and directed along the binormals $\mathbf{b}(S,t)$

$$V = \frac{b}{4\pi} x(S,t) b(S,t) A(L,d), \qquad (6)$$

where L is the length, d is the transverse dimension of the deforming region, S is the current value of length, and b is the absolute value of the volume incompatibility of the displacement field (Burgers vector). The scale factor $A = \ln\frac{2L}{d} - 1$ for $L \gg d$ and $A = 0.2$ for $L \sim d$.

A transfer in the equation of motion for plastic flow to a non-linear equation allows us to calculate the curvature $x(S,t) = 4\beta\,\text{sech}\,2\beta(S-\alpha t)$, where α and β are the torsion and maximum curvature of the deforming region, respectively. It should be noted that β prescribes the value of plastic flow and α – the rate of its variation along S.

Having selected the global coordinate system for the z-axis to be directed along L and X and E to be varied within the limits of the transverse dimensions, in the Cartesian coordinate system for the strain $\varepsilon_{\mu\nu}$ and rotation $\omega_{\mu\nu}$ tensors we have

$$\varepsilon_{zz} = 1 - \frac{2}{1-V^2}\text{sech}^2\xi, \qquad (7.1)$$

$$\varepsilon_{zx}(\omega_{zx}) = \frac{1}{1+\nu^2}\left\{\frac{\alpha b x t}{\pi(x^2+y^2)}\left(\frac{1}{ch^2\xi}-\frac{1}{ch^2\eta}\right)\pm\left(\frac{\sin\nu\xi\,sh\xi}{ch^2\xi}-\frac{\nu\cos\nu\xi}{ch\xi}\right)\right\} \qquad (7.2)$$

$$\varepsilon_{zy}(\omega_{zy}) = \frac{1}{1+\nu^2}\left\{\frac{\alpha b y t}{\pi(x^2+y^2)}\left(\frac{1}{ch^2\xi}-\frac{1}{ch^2\eta}\right)\pm\left(\frac{\cos\nu\xi\,sh\xi}{ch^2\xi}-\frac{\nu\sin\nu\xi}{ch\xi}\right)\right\} \qquad (7.3)$$

The subscripts in brackets in the left-hand part of (7) correspond to the rotations $\omega_{\mu\nu}$, $\xi = 2\beta z + \dfrac{2\alpha\beta b}{\pi} At$, $\eta = \xi - 2\beta z$, and $\nu = -\dfrac{\alpha}{\beta}$.

Quantity $\dfrac{\alpha\beta A}{\pi}$ represents the velocity of wave propagation along the region – the velocity of plastic flow variation in the global coordinate system. In the local coordinate system the velocity is determined by torsion α only.

It should be noted that initially the scale factor A was included into the determination of time t in the common system of coordinates. This might imply that time during plastic deformation has no absolute meaning but is a parameter characterizing the process, and velocity is controlled by torsion only.

The evolution of the axis of a plastically deformed region and strain rate reproduced using formulas (6), (7) are given in Fig. 1. The arrows indicate the values and direction of the velocity polarized along binormals **b**, perpendicular to direction **t** of propagation of the shape variation wave.

The length of this wave is $\lambda = \dfrac{\pi}{\alpha}$, and the 'wave number' $k = 2\alpha$. Its frequency possesses square dispersion and is equal to $\omega = \dfrac{2\alpha^2}{\pi} A$. In other words, wave characteristics of plastic waves are controlled by torsion (or by torsion and dimensions of the region).

The maximum variation of the velocity occurs in the sites of the helical curve bending. The variation in the direction of deformation velocity along the curve is due to the rotation of the flow vector and the change in the curvature. In the case of a completed rotation, the wave polarization plane rotates by the angle $\Omega = \alpha L$, i.e., the velocity obeys the law of parallel transfer [6,7].

The dependence of the shape of the region and deformation velocity on β (Fig.2) indicates that with increasing β the shape sharply changes as do the velocity value and direction. In the sites with large current curvature and bending, the velocity changes its sign to the opposite, which might give rise to discontinuities in the deforming region.

A similar deformation behavior is observed under condition of decreasing torsion α (Fig.3). Torsion controls the wavelength and velocity of plastic flow propagation along the region (rate of curvature variation). In the case of small values of α – large wavelengths, the shape of the region can undergo critical changes until its separation into parts. In the case of large values of α – small wavelengths, the shape would change gradually at a nearly constant strain rate along the entire length. The only change is that in the (flow) velocity direction, rotating with the binormal vectors.

Figure 4 demonstrates the shape of the region and deformation velocity as a function of the region length. At low values of L, the velocity is nearly homogeneous, while in the case of a longer region the situation is different. In other words, fragmentation of the deforming region is the result of a tendency of deformation and its velocity towards its homogeneity.

Thus, at large α and small β, L, plastic deformation is practically homogeneous throughout the entire deforming region. On the contrary, small α and large β, L (large lengths and amplitudes of non-linear plastic waves) are responsible for sharply inhomogeneous localized deformation until its critical behavior and refinement of the material. The mechanism of this refinement will be addressed in the discussion of fragmentation.

The dynamics of plastic deformation itself represents the motion of a 'fold' and the region of its localization. Figures 5, 6 present the calculated evolution scenarios of the shear deformation fold motion along axes z and x. Deformation $\varepsilon_{xz}$ along the z-axis, tending towards the edge of the region, becomes constant and equal to $\left(\dfrac{\alpha\beta}{\alpha^2+\beta^2}\right)$. Along the x-axis, the value of $\varepsilon_{xz}$ increases in time towards the edge, but for large times it is proportional to $\dfrac{\beta^2}{\alpha^2+\beta^2}\dfrac{L}{d}$, i.e., it also depends on the scale of the region. Under critical conditions, at small α and large β, the value of $\varepsilon_{xz} \sim \dfrac{\alpha}{\beta}$ is small along the z-axis, while the value of $\left(\varepsilon_{xz} \sim \dfrac{L}{d}\right)$ is small along the x-axis.

The calculation of the compression-tension deformation $\varepsilon_{zz}$ (Fig.7) reveals compression at small values of α and z, and tension with increasing z, which closer to the edge of the region ($z \to L$) tends to be constant. At large values of α and small values of β deformation $\varepsilon_{zz}$ becomes tensile for any lengths of the region.

The evolution discussed herein represents a single act of plastic flow, whose recurrence is supported by the applied stresses. The flows of displacement (defect) field discontinuities and energy from the regions of localization, which change the entropy production and provide the locomotion of the process, make it possible to 'freeze' the resulting shape variation.

## 3. Constitutive equation

In the course of localized superplastic deformation, defect and energy flows are emitted from the deforming region, concurrently with the heat and mass transfer. The work function thus performed determines the specific entropy production S (dissipative function) and the flux of entropy, I

$$S = \dfrac{(Q_i J_i \mu)\partial_i T}{T^2} - \dfrac{J_i \partial_i \mu}{T} - \dfrac{\partial_i T \omega_{ik} \upsilon_k}{T^2} + \dfrac{\sigma_k \upsilon_k}{T}$$
$$I_i = \dfrac{\omega_{ik}\upsilon_k}{T} - \dfrac{Q_i}{T} - \dfrac{\mu J_i}{T},$$
(8)

where $\omega_{ik} = \begin{cases} \varepsilon_{ijk}\alpha_j - \text{ at the microscale; } \alpha \text{ -dislocation density,} \\ (\partial_k U_i - \partial_i U_k) - \text{ at the macroscale - antisymmetric part} \\ \qquad\qquad \text{of plastic distortion (rotational mode),} \end{cases}$

υ is the velocity of plastic deformation, σ are the local stresses, T is the temperature in the locally equilibrium conditions, $\varepsilon_{ijk}$ is the Levi-Civita symbol, $Q_i = -\dfrac{\kappa \partial_i T}{\rho}$ is the specific heat flux, $\kappa$ is the heat conduction, ρ is the material density, $\partial_i T$ are the temperature gradient components, $J_i = -\dfrac{D\partial_i c(z)}{\rho}$ is the specific mass flow, D is the diffusion coefficient, $\partial_i c$ are the concentration gradient components, and $\mu = \mu_1 - \mu_2$ is the difference between the chemical potentials of the material and the diffusing component.

The first and second terms in the right-hand part of the equation for S (8) are the production of entropy due to heat release and mass transfer; the other two are due to plastic deformation [6,7]. The former two terms are of the second order infinitesimal with respect to the gradients: $\delta_i T \delta_i T$, $\delta_i T \delta_i C$, $\delta_i C \delta_i \mu$, while the thermoplastic term is of the first order infinitesimal.

The second law of thermodynamics gives us the constitutive material equation – the equation of state for superplasticity of a deforming system

$$\sigma_k \upsilon_k - \frac{(Q_i - J_i \mu)\partial_i T}{T} - J_i \partial \mu - \frac{\partial_i T}{T} \omega_{ik} \upsilon_k \qquad (9)$$

It is now evident that the process of superplastic flow supporting this state becomes irreversible. In the case of smallness of the second-order terms with respect to gradients, relation (9) is given by the following:

$$\left(\sigma_k - \frac{\partial_i T}{T} \omega_{ik}\right) \upsilon_k \geq 0 \qquad (10)$$

Similar to the case where in an elastic state its characteristic – deformation, is related to the stresses by the Hooke law, under plastic and superplastic conditions their characteristics – macroflows and densities (orientation and rotation modes), are related by the local stresses via expressions (9) and (10). The deviator $\sigma_k$ in (9), (10) is the irreversible stress equal to the difference in the total local stresses and their values at the yield point (elasticity limit).

The inequalities (9), (10) correspond to SP and plasticity with reinforcement, an equality being possible for the case of ideal plasticity (SP), at $\upsilon_k \neq 0$. In the absence of a temperature gradient, (10) coincides with the equation of state well-known in the mechanics of plasticity [12]. The discussion of general properties of the constitutive equation and the dissipation function could be found in [12]. Let us dwell on some of the conclusions associated with a particular form of the irreversible force:

- relations (9), (10) suggest that an increment in the work function (power) is positive – Drueker's postulate;
- during unloading ($\sigma<0$), equality (10) is satisfied only for $\upsilon_k=0$. This implies that unloading is performed via elastic deformation;
- inequality (10) during unloading is satisfied only for $\upsilon_k<0$ – inverse plastic deformation that takes place at the total stresses lower than the yield stress (elasticity limit), i.e., is accompanied by the softening Bauschinger effect;
- ideal thermoplastic state without reinforcement, which is determined by equality (10), corresponds to a steady state ($\frac{\partial S}{\partial \upsilon_k}=0$)

$$\sigma_k - \frac{\partial_i T}{T} \omega_{ik} = 0$$

and the total thermodynamic flow is zero. This implies that every act of plastic deformation has a corresponding state at which stress concentrators are compensated by the rotational mode, given a temperature gradient. This compensation is the more pronounced the lower is the specimen temperature. Hence, the σ-ε curve would have jumps in stresses after propagation of every plastic wave, which would increase with deformation. It is these oscillations – the Portevin-Le Chatelier effect (PLC) – which are observed in the experimental curves (Fig. 8 from [13]). The variation in Δσ from jump to jump corresponds to a variation in the material rotation. In the case of superplastic deformation, it follows from the equation of state that

$$\left(\sigma_k - \frac{\partial_i T}{T}\omega_{ik}\right)\upsilon_k \geq \frac{(Q_i - J_i\mu)}{T}\partial_i T - J_i\partial_i\mu;$$

- an increase in the power of plastic deformation has to be not merely positive, it has to be larger than the increment due to the heat and mass transfer – a generalized Druker stability postulate;
- another steady-state condition for plastic flow is achieved during transformation of the absorbed heat and mass transfer (migration and diffusion) into the density of the mechanical energy flux

$$-(Q_i - J_i\mu) = \omega_{ik}\upsilon_k;$$

- without variation in the level of stresses. This state, corresponding to plastic flow and rotations activated by heat and mass transfer at low stresses, represents a special plastic deformation mechanism – creep;
- still other state follows from (9) at high temperatures, where

$$\sigma_k\upsilon_k \geq J_k\partial_k\mu$$

at $J_k = -\frac{D\delta_k C}{\rho}$ and $D_k\mu = l^3\sigma_k$, we have

$$\upsilon_k \leq Dl^3\partial_k c$$

- an equation of state for diffusion-induced creep. The velocity of creep is independent of the level of stresses and is controlled by the diffusion coefficient, diffusing volume, and quantity $\partial_k c = \partial_k \delta c(z)$ – the gradient of concentration fluctuations;
- the constitutive equation of 'pure' plasticity (special case (10)) with all its corollaries, which is obtained here relying on the non-equilibrium dynamics and physics of defects, coincides with the constitutive equation in the mechanics of plasticity [12], which indicates the identity of both approaches to treating plasticity.

Relation (9) is valid not only under condition of plastic deformation, but also for the evolutionary processes in any inhomogeneous medium containing displacement field discontinuities: in gases, fluids, and solids. It should be noted that plastic flow can result not only from mechanical stresses but also heat and mass flows and material rotations. Given a temperature gradient of a low local density (air) and low temperature, the heat released (or absorbed) by the region generates a vortex, $\omega_{ik}$, moving at velocity $\upsilon_k$ and unloading stresses ($-\sigma_k$), directed inside the vortex (or loading stresses directed outside). All these features are typical for different kinds of tornadoes and hurricanes. In other words, in this particular case the constitutive equation is the equation of state of a whirlwind.

Figure 9 presents a calculated helical wave of shape variation – the area of the whirlwind at V directed along the y-axis, and $\omega_{xy}=\omega_z$ corresponds to rotation along the z-axis in the Cartesian coordinates in accordance with the equation of state.

## 4. Concentration fluctuations under conditions of plastic deformation

Let us address the fluctuations of concentration, δc, in the deforming region at Φ≠0 under conditions of plastic equilibrium (see equation (3)).

As before [14], let us write free energy for the fluctuations of concentration

$$F = \int\left\{-a|bc|^2 + \frac{b}{2}|\delta c|^4 + \left|\left(i\partial\mu + \frac{1}{l^2}\varepsilon_{\mu\upsilon k}\Phi_\upsilon r_k\right)\delta c\right|^2 + \Phi^2\right\}d\upsilon. \qquad (11)$$

Let us assume the variations in concentration to be maximal along the z-axis only, i.e., $\frac{\partial c}{\partial x}, \frac{\partial c}{\partial y} \ll \frac{\partial c}{\partial z}$ and c=c(z). The fluctuations, $\delta c$, are short-wave ones and, in the general case, coefficient $\alpha = \alpha(T) + a'(q - q_0)^2$ [14] for $|q - q_0| \ll q_0$. Vector $\vec{q}_0$ in the reciprocal space corresponds to vector $\vec{l}$ in the straight space, which limits the scale of plastic misorientation Φ, so that $\vec{q}_0 \vec{l} = 2\pi$. The term in α, depending on $\vec{q}$, is responsible for the formation of segregations (weak crystallization) in the region of concentration variations. Since segregations do not appreciably affect superplasticity, so we can omit $\alpha'|q - q_0|$ and assume coefficient α to depend on temperature only $\alpha(T) = \tilde{\alpha}(T - T^*)$, where $T^*$ is the temperature of the loss of stability of the high-temperature phase.

Minimizing (11) with respect to δc, we obtain

$$\frac{\partial^2 \delta c}{\partial^2 z} - \frac{\partial \delta c}{\partial t} + \delta c - |\delta c|^2 \delta c = 0, \tag{12}$$

where $\frac{\partial \delta c}{\partial t} = \upsilon \frac{\partial \delta c}{\partial z}$, and quantity $\upsilon = \frac{2u_z}{al^2}$ plays the role of the velocity of fluctuation motion along the z-axis, which depends on plastic displacements $u_\mu = \varepsilon_{\mu\upsilon k} \Phi_\upsilon r_k$, temperature (via α(T)), and the scale of misorientation.

In order to derive an equation, relating concentration fluctuations and plastic flow, let us minimize (11) with respect to $u_\mu$, representing parameter $\delta C = |\delta C| e^{i\theta(r)}$ as a product of the absolute value and the argument, wherein the inhomogeneous phase of fluctuations would determine the total velocity of superplastic flow.

The resulting equation together with (3) would form the following system:

$$\partial_\mu \theta = \xi \upsilon_\mu + \frac{\lambda}{|\delta c|^2} \varepsilon_{\mu\upsilon k} \partial_\upsilon \Phi_k \tag{13}$$

$$\varepsilon_{\mu\upsilon k} \partial_\upsilon \Phi_k = V_\mu + c^\beta_{\mu\upsilon} p^\upsilon_\beta, \tag{3}$$

where $\lambda = \frac{b}{a}, \xi = \frac{al^2}{2}, |\delta c| \neq 0$.

These two equations describe the relationship between long-wave plastic variations and short-wave concentration variations, exchanging energy with each other.

An incompatibility between (13) and (3) would result in separation (fragmentation) of the deforming region.

This energy exchange would be most efficient under conditions of a resonance, where velocities $\upsilon_\mu$ and $V_\mu$ are of the same order. In this case, shape variation would occur by a collective motion of the area of the size of a concentration kink – i.e., it would follow the solution of (12). This solution is given by

$$\delta c \sim \pm th \frac{1}{2} \frac{z - \upsilon t}{\sqrt{1 - \upsilon^2}}, \tag{14}$$

where υ and t are determined after solving (12).

Combining (3) and (13), we obtain the velocity of superplastic flow

$$\partial_\mu \theta = \frac{\lambda}{|\delta c|^2} V_\mu + \xi \upsilon_\mu + \frac{\lambda}{|\delta c|^2} c^\beta_{\mu\upsilon} p^\upsilon_\beta. \tag{15}$$

The first term in the right-hand part of (15) represents the above-discussed velocity of plastic deformation due to the wave process (expression (6)) increased by a factor of $\frac{\lambda}{|\delta c|^2}$.

It should be underlined that during calculation of $V_\mu$ we had already included the rotational part of plastic distortion ($\Phi`$) and plastic displacement ($u_\mu$). Thus, in the two subsequent terms in the right-hand part of (15), the first term ($\xi \upsilon_\mu$) is determined by the gradient of external sources, and the second – by the 'delayed' polarization, $V_\mu$, which begins to act when it reaches its maximum value. Note however that in the cases where in the deforming regions the moduli $c^\beta_{\mu\upsilon} \to 0$, then this term would determine intragaranualr deformation (slip) only.

Let us now address the term corresponding to the outside sources.

## 5. Creep and superplasticity

If an outside source is the gradient of the chemical potential $\partial_\upsilon \mu$, then $\upsilon$ would be the diffusion velocity prescribed by the following relation:

$$\upsilon_\nu = -\frac{D}{\rho T}\partial_\nu \mu,$$

where D is the diffusion coefficient, $\rho$ is the material density, and $T$ is the temperature.

Let us average (15) over the boundary contour, denote $\frac{1}{L}\oint \langle \partial_\mu \theta \rangle \frac{dl}{\xi} = \langle \dot\varepsilon \rangle$, and express the difference in chemical potentials via the external stresses

$$(\mu_2 - \mu_1) = l^3 \sigma_{ext},$$

where $l$ is the scaling factor (at the microscale, $l^3$ is the atomic volume). Then

$$\langle \dot\varepsilon \rangle = \frac{Dl^3 \sigma_{ext}}{\rho T L}. \tag{16}$$

Dividing (16) by $L^2$, we obtain the velocity of the Cobble creep and dividing the boundary by the length, we obtain the Herringer-Nabarro formulas: $\langle \dot\varepsilon \rangle = \frac{Dl^3 \sigma_{ext}}{\rho T L^3}$. These formulas are valid at all scale levels differing in the values of $L$ and $l$.

The displacements (deformations) controlled by the diffusion mechanism would be given by

$$\langle u \rangle = \frac{\tilde{a}\left(1-\frac{T}{T^*}\right)Dl_0^2 l^3 \sigma_{ext}}{2\rho L^{3(2)}}. \tag{17}$$

They strongly depend on the scale factors, $L$ and $l$, and are sensitive to the transformation temperature. At $T \to T^*$, the fluctuations participating in the decomposition are unlikely to take part in the formation of plastic displacements.

When the diffusion coefficient is controlled by the temperature, $D = D_0 e^{-\frac{Q}{T}}$, the maximum plastic deformation is achieved at $T_m = \frac{QT^*}{QTT^*}$. Since $Q = \frac{T^*}{n}(n>0)$, then $T_m = \frac{T^*}{(n+1)}$ is a $\frac{1}{n+1}$-th fraction of $T^*$. The larger is Q(n<1), the closer is $T$ to $T^*$.

For $Q \sim T^*, T_m \sim 2$, the smaller is Q(n>1), the lower is $T_m$. In other words, the higher is the energy of diffusion activation, the closer to the stability loss threshold is the maximum plastic deformation and, vice versa, at low energies the maximum deformation is achieved at low temperatures.

Under the condition of plastic equilibrium (15), we have to differentiate between two cases:

1. Without any transformations inside the deforming area, the fluctuation phase is homogeneous and $\partial_\mu \theta = 0$. In this case, diffusion stimulates plastic flow (variation in curvature along the boundary): $\dfrac{V_\mu}{a|\delta c|^2}$ – creep and a possible subsequent intragranular slip (equation (15) at $\partial_\mu \theta = 0$). Substituting into $\dfrac{V_\mu}{|\delta c|^2}$ the value of $V_\mu$ from (6) and $|\delta c|^2$ from (14), we obtain the dependence of deformation on time. This dependence, $E_{zz}(t)$, is given in Fig. 10, where curve 1 corresponds to the initial deformation and curve 2 corresponds to the deformation strengthened by the fluctuations in concentration. Curve 1 has two stages of strain development – an increase and a dip with a maximum corresponding to time $t_1 = \dfrac{L}{\alpha}$. Curve 2 has three stages: the first stage lasting until time $t_1$, the second – until time $t_2 = \dfrac{L}{\upsilon}$ ($\upsilon$ is the diffusion velocity), and the third, an increase in deformation taking place after $t_2$ up to a break in the curve – fracture. This completely coincides with the stages of a classical experimental curve *E(t)* under creep.

Thus, two velocities control the rate of plastic flow variation along the region: the velocity of a shape (curvature) soliton, α, and the diffusion rate – a kink in the concentration fluctuation curve, V. It should be underlined that the governing parameter is the rate of variation in the plastic deformation velocity (α) rather than the deformation velocity itself. The stage-like character of the E(t) curve is due to these two parameters. The closer are |α| and |V|, the longer would be stage II in the E(t) curve (which corresponds to the maximum energy exchange between the two processes). Diffusion in this case is not only a source of deformation, but, which is no less important; it ensures an enhancement of the plastic flow. This becomes possible, given a definite relationship between these parameters $L, \alpha, \upsilon, T, \rho, D, \sigma_{ext}$. A violation of these relationships would result in either no enhancement, or a deformation jump – fracture.

An approximation of the temperature towards $T^*$ would additionally enhance the wave deformation process via $\dfrac{1}{a(T-T^*)}$, despite the decreasing of diffusion-controlled plastic displacements. On the other hand, the latter might give rise to incompatibility of equations (3) and (13), in other words, $\varepsilon_{\mu\upsilon k} \Phi_\upsilon r_k$ in (3) and (13) would not coincide. In this case, the deforming region would get fragmented even at small strains; this would give rise to new outside sources of yielding, which will be addressed in the next section.

2. In the presence of transformations in the deforming area, the phase of concentration fluctuations is inhomogeneous, $\partial_\mu \theta \neq 0$. It is this phase which determines the total velocity of superplastic deformation. Diffusion makes an independent contribution into superplasticity and, which is even more important, enhances the other contributions. The above considerations for creep are valid for this case as well.

Thus, the presence of transformation is a characteristic and necessary feature of superplastic flow. Nevertheless, it is impossible to present superplasticity as a phase transition (superfluidity) due to the leading role of the plastic dynamic equilibrium. Note one more phenomenon associated with diffusion. If the phase of concentration fluctuations depends on time, it will be easy to demonstrate that $\frac{\partial \theta}{\partial t} = \mu_2 - \mu_1 = l^3 \sigma$. In other words, phase variation in time determines the jump of the chemical potential in certain 'narrow' sites and generates elastic stresses $\sigma$, which could serve a source of plastic flow instead of the external stresses in accordance with the above-described scenario.

The concentration-based mechanism of superplasticity is hardly different from the structure-based one, according to which the role of concentration is played by the order parameters of the structural transition [6,7].

## 6. Fragmentation and grain-boundary sliding

An independent mechanism of plastic deformation taking place at grain boundaries is their fragmentation, misorientation of the structural elements, and grain-boundary sliding (GBS) due to mass transfer. In the initial stages of SP, the flow is associated with this particular mechanism.

According to the above discussion, fragmentation is the mechanism of achieving plastic equilibrium; it develops as a result of the fact that the perfect plastic equilibrium – compatibility of (3) and (13) at large lengths of the boundaries, is not virtually fulfilled; this implies that (13) is compatible with an equation of the type of Eq. (3): $\varepsilon_{\mu\nu k} \Phi_\nu r_k = V_\mu + c^\beta_{\mu\nu} p^\nu_\beta + F_\mu$, containing certain outside sources, $F_\mu$, in the right-hand part, which ensure this compatibility via fragmentation, GBS, etc.

The mechanism of fragmentation might consist in the following (see Figs. 2, 3). At large β and small α, in the sites of the helical curve bending, large-value plastic flows have opposite directions. In the convex portion, they meet halfway, while in the concave portion they are directed oppositely (Figs. 2, 3). This gives rise to the development of dense and loose areas in the material. It is in the latter areas where discontinuities develop, which form several shorter-length deforming regions. Within these areas, the plastic flow J=0 and the distance to one of the discontinuities corresponds to the plastic wave length $\lambda = \frac{\pi}{2}$. Figure 4 presents the deformations, and the inset – the deformation velocities for different lengths of the regions. At small L, the plastic deformation velocity is practically the same along the entire length. In other words, the plastic flow motion is homogeneous. As the length is increased, this homogeneity disappears. Therefore, fragmentation of the deforming region is a tendency of the deformation and the flow towards uniformity. Condition J=0, the condition of fragmentation, is satisfied in the sites of bending, where curvature is not accommodated by the flow.

Given outside sources $\partial_\mu f$ of the diffusion type, condition J=0 instead of fragmentation converts the boundary into a 'mantle' due to the transfer of linear defects into it. The width of the mantle is determined by the rotation [6,7]

$$d = Le^{\frac{-4\pi(\nabla f \vec{b})}{xb}}. \quad (18)$$

Both fragmentation and formation of the mantle during diffusion controlled creep take place under the conditions of suppressed localized plastic flow, J=0. These processes result in the constraint of deformation and its possible accommodation of the rotation of the adjacent grains as a single mechanism of energy dissipation.

It is easy to estimate the strain rate during grain rotation – grain-boundary sliding, and mass transfer towards one of the cones of the region.

During the mass transfer M at an angular velocity ω, the grain of mass M rotates due to the law of conservation of angular momentum and moves in the opposite direction at an angular velocity Ω. Since the values of these velocities are small, so $\omega \sim \frac{\varphi}{t_n}$, where $t_n$ is the time of measurements, $\varphi = \frac{l}{R}$ is the angle of displacement of mass m by the distance $l$, and R is the 'average' grain radius. From the law of conservation of angular momentum for the linear velocity $\upsilon_g = \Omega R$ during grain rotation, we obtain

$$\upsilon_g = \frac{\Delta V \rho l}{V \rho_g t_n}, \qquad (19)$$

where V is the grain volume, $\rho$ is the boundary material density, $\rho_g$ is the grain density, and ΔV is the displaced material volume. In the case of curvature displacement along L, the rate of GBS is equal to $\dot{\varepsilon} = \frac{\Delta V \rho}{V \rho_g t_n}$, i.e., it is determined by the respective mass ratio. Assuming the deforming volume to be a cylinder with the cross-section area $\pi d^2$ and length $\alpha b (\ln \frac{2L}{d} - 1) t_n$, where $\alpha b (\ln \frac{2L}{d} - 1)$ represents the velocity of curvature displacement along L, we obtain

$$\dot{\varepsilon} = \frac{\pi d^2 \alpha b \rho \left( \ln \frac{2L}{d} - 1 \right)}{V \rho_g} \qquad (20)$$

Having selected elastic torsion $\alpha = \frac{\sigma}{Gd}$, σ is the external shear stress, G is the shear module let us take the width, d, in the form given by (18), where $\nabla f = \frac{D}{\rho kT} \left( \nabla \mu \vec{b} \right)$ is the diffuse outside sources, $\nabla \mu$ is the chemical potential gradient, D is the diffusion coefficient, and $\left( \nabla \mu \vec{b} \right) = l^3 \sigma$, and, substituting these into (20) for the strain rate of GBS, we obtain

$$\dot{\varepsilon} = \frac{\pi^2 L l^3 D}{V \rho_g \beta G k T} \sigma^2 e^{-\frac{\pi D l^3 \sigma}{\rho \beta b k T}}. \qquad (21)$$

Within the context of superplasticity, GBS is a thermo-elastically activated plastic process stimulated by the concentration and structure fluctuations. The role of activation energy here belongs to quantity $\frac{\pi D}{\beta b \rho} \left( \nabla \mu \vec{b} \right)$. The rotation of the binormal vector $\vec{b}$ during its motion along the deforming area under conditions of flow variation makes this quantity a time-variable parameter. In other words, GBS occurs in a kinetic mode, despite the thermo-elastic activation. The dependence of $\dot{\varepsilon}$ on stresses has a maximum in $\sigma = \sigma_0 = \frac{2 \beta \rho b k T}{\pi D L^3}$, which is equal to

$$\dot{\varepsilon}_{max}(\sigma_0) = \frac{2\beta\rho^2 b^2 kT}{3\pi\rho_g VGL^2 D} \qquad (22)$$

The maximum velocity increases with an increase in temperature, density, curvature, and flow vector incompatibility. The physical state of the region favoring GBS is characterized by low shear moduli, small grain volume and low grain density, short diffusion lengths, and low diffusivity. Note that an increase in $\dot{\varepsilon}_{max}$ is favored by an increased density in the material of the boundaries rather than their 'loosening'.

Let us analyze the dependence of $\dot{\varepsilon}$ on grain size. Assume $L = nR$, where $0 < n < 2\pi$ is the radius fraction R of the grain in the length L, $\beta \sim \frac{1}{R}$ is the maximum curvature, and $V_g$ is the constant. Then $\dot{\varepsilon} = AR^2 e^{-BR}$, where $A = \frac{\pi^2 l^3 D\sigma}{V\rho_g GkT}$ and $B = \frac{\pi l^3 D\sigma}{\rho b kT}$. Curve $\dot{\varepsilon}(R)$ has a maximum at $R = R_0 = \frac{2\rho b kT}{\pi D l^3 \sigma}$, equal to

$$\dot{\varepsilon}(R_0) = \frac{\rho^2 b^2 kT}{2V\rho_g G l^3 D} \qquad (23)$$

This nonmonotonic behavior is due to a macrodependence $\dot{\varepsilon}(R)$, rather than variation in the microscopic diffusion coefficient D with R, as it is commonly assumed [15]. Optimal grain radius $R_0$ increases with temperature, (which is consistent with the experiment [15]), grain-boundary density, and the Burgers vector flow, but decreases with increasing stresses.

The dependence of $\dot{\varepsilon}$ on the temperature has a maximum at the temperature $T_0$ equal to the multiplier before $1/T$ in the exponent (21)

$$\dot{\varepsilon}(T_0) = \frac{\pi\rho b\sigma}{V\rho_g G}. \qquad (24)$$

The process of GBS itself, whose velocity is determined by formula (21), consists in the following. Under external loading, the curvature of the deforming region (boundary) gives rise to the development of plastic flow J, proportional to the Burgers vector curvature and flow, directed transverse the curved boundary. This curvature (shape) soliton also propagates along the region at a velocity determined by the incompatibility vector torsion, length and flow. In the initial stages, for small values of torsion – low velocities, there is condition J=0 for mechanical fragmentation of the deforming area in the sites of the plastic wave minima. This condition J=0 is also fulfilled under the action of outside sources – diffusion, forming the 'mantle' of the aforementioned length, changing the grain shape and size. Both fragmentation and diffusion-controlled creep result from the conditions hindering the localized plastic flow rather than its reason. In the case of creep, the reason is compensation of the plastic curvature X by the outside sources and stagnation of the flow across the region of localization. In the case of fragmentation, this is stagnation of the flow in the sites of bending of the deforming region – plastic wave minima in the case of small torsion values α→0, i.e., in the case of retarded motion of the flow along the region.

Under constrained conditions, shape variation is accompanied by mass transfer, accommodated by rotation of the adjacent grains, which is the case of GBS.

## 8. Summary

It has been shown that the wave plastic process underlies all plastic deformation phenomena; it makes an independent contribution into them, allows describing them from the same perspective and makes it possible to investigate the relationship between the physics of defects and the mechanics of plasticity. The defects are not limited to microdefects – dislocations and disclinations. The meso- and macrodefects, such as displacement vector field discontinuities, representing a non-linear process of localized plastic deformation, also play their independent roles.

The wave characteristics of plastic flow are determined by the geometric (curvature, torsion), scale (length, transverse dimensions), and physical (elasticity moduli, density, sound propagation velocity) parameters of the deforming areas of the material. The length of a non-linear wave $\lambda = \dfrac{\pi}{\alpha}$ is controlled by torsion and its frequency possesses square dispersion $\omega \sim \dfrac{2\alpha^2}{\pi}$. At small lengths (large α) and low curvatures β, the localized plastic deformation is practically homogeneous; on the contrary, small values of α and large β give rise to sharply inhomogeneous deformation.

The constitutive material equation – the equation of superplastic state, obtained in this work relying on the concepts of non-equilibrium thermodynamics and description of defects from the standpoints of the wave theory, relates the velocity of plastic deformation, rotational modes, local irreversible stresses, temperature, heat and mass flow densities and, in the special case of 'pure' plasticity, it coincides with the constitutive equation of the mechanics of plasticity.

The corollaries to this equation are the generalized Drucker postulate and its general form, the Bauschinger effect. Its stationary states determine the Portevin–Le Chatelier effect, creep, and diffusion-controlled creep.

The coincidence, in a particular case, of the constitutive equation of plasticity with that of the mechanics of plasticity implies an identity of both approaches.

In the case of less continuous media, the constitutive equation represents the equation of state of a hurricane (whirlwind).

The superplastic state is a state of dynamic plastic equilibrium accompanied by the concentration (structure) fluctuations – transformation in the deforming region (boundary).

The expression for velocity of SP flow, which is controlled by this equilibrium, contains three terms: the first is the velocity due to the wave process in the boundary, increased by a factor of $\lambda / |\delta c|^2$, the second is prescribed by the gradient of outside sources (diffusion and the like), and the third is the delayed polarization that, given a transformation, represents intragranular slip.

The transformation or decomposition is a characteristic and necessary feature of superplastic flow. In the absence of this peculiarity, plastic deformation occurs via creep.

If the outside source in the second term is a gradient of the chemical potential, then the plastic deformation velocity, averaged with respect to the boundary contour and divided by the squared lengths of the deforming region, would be the velocity of diffusion-controlled Cobble creep and, when divided by the length, would represent the Nabarro-Herring velocity. Deformation under diffusion-controlled creep strongly depends on the size of the region, diffusing volume, and is sensitive to the transformation temperature. Two cases could be distinguished in the aforementioned transformation:

1. The phase of fluctuations is homogeneous, there is no transformation. Diffusion stimulates creep. The curve of dependence of tensile deformation on time has three stages. The first stage lasts until $t_1 = \frac{L}{d}$, the second – until $t_2 = \frac{L}{\upsilon}$, and the third begins after $t_2$, an increase in deformation until discontinuity in the curve – fracture. The processed is controlled by the velocity of curvature variation along the boundary (α) and the diffusion velocity ($\upsilon$).

2. The inhomogeneous phase of fluctuations determines the total SP flow velocity (all three terms were discussed above).

Thus, in this study we have discussed fragmentation and GBS, their mechanisms and peculiar features. We have obtained formulas for deformation velocity under GBS and analyzed the reasons for its non-monotonic dependence on grain size, applied stresses, and temperature.

**FIGURE CAPTIONS**

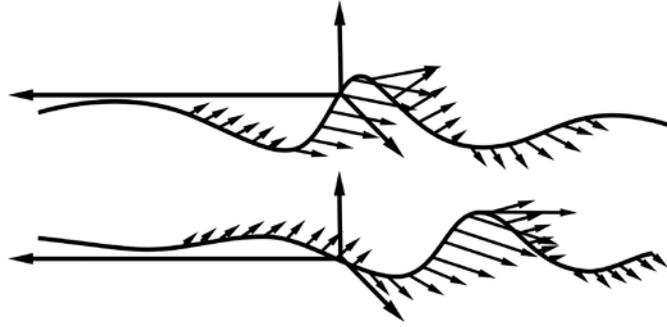

**Fig.1.** Variation in the shape of the deforming region and plastic deformation velocity (arrows).

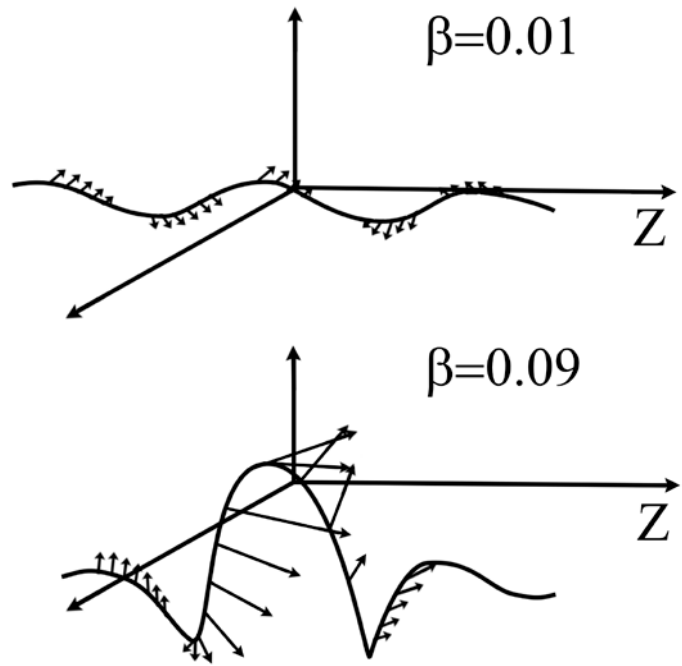

**Fig.2.** Dependence of the shape of the deforming region and plastic deformation velocity on curvature β.

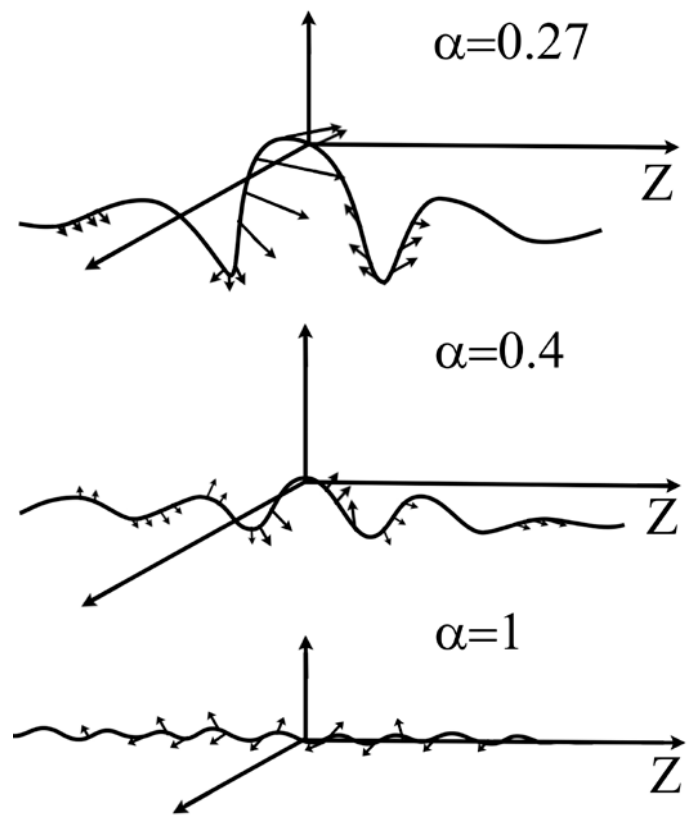

**Fig.3.** Dependence of the shape variation on torsion α.

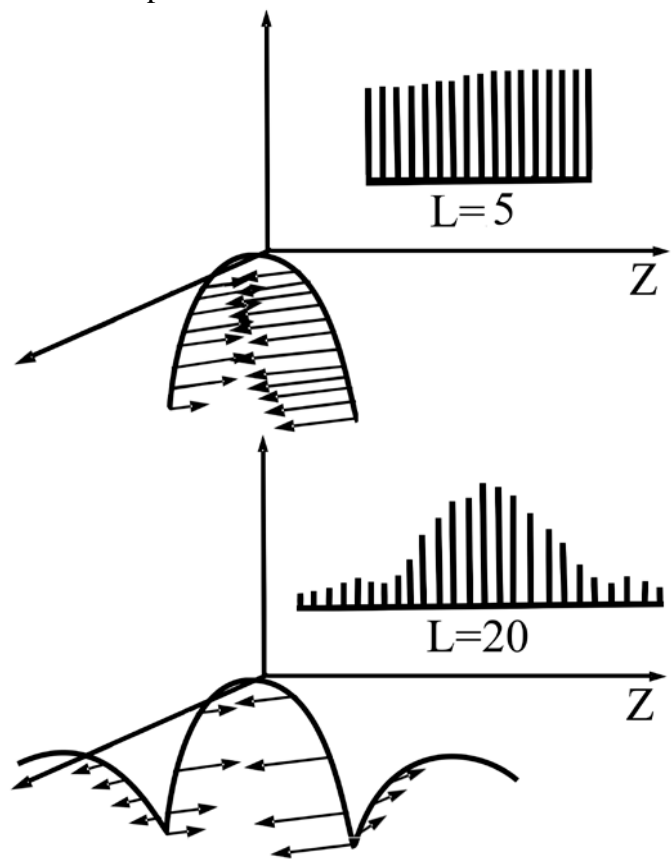

**Fig.4.** Shape of the deforming region and deformation velocity (insert) as a function of the deforming-region length L.

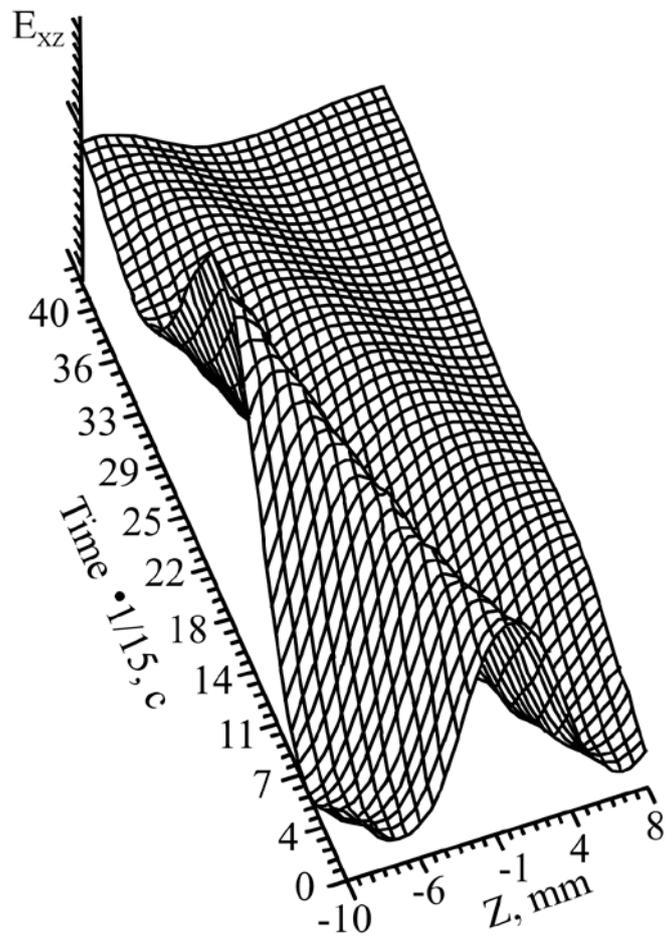

**Fig. 5.** Evolution of $E_{xz}$ along z.

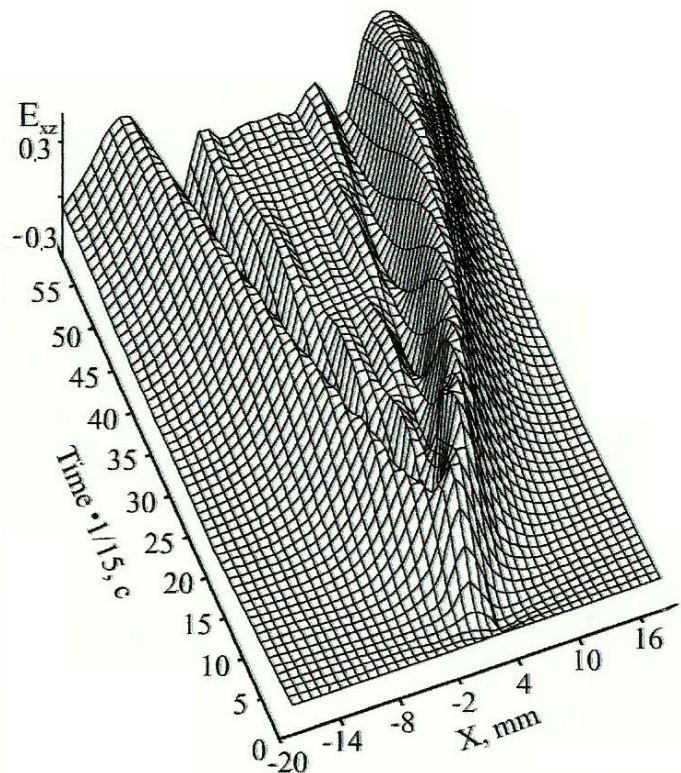

**Fig. 6.** Evolution of the 'fold' of local displacements $E_{xz}$ along x.

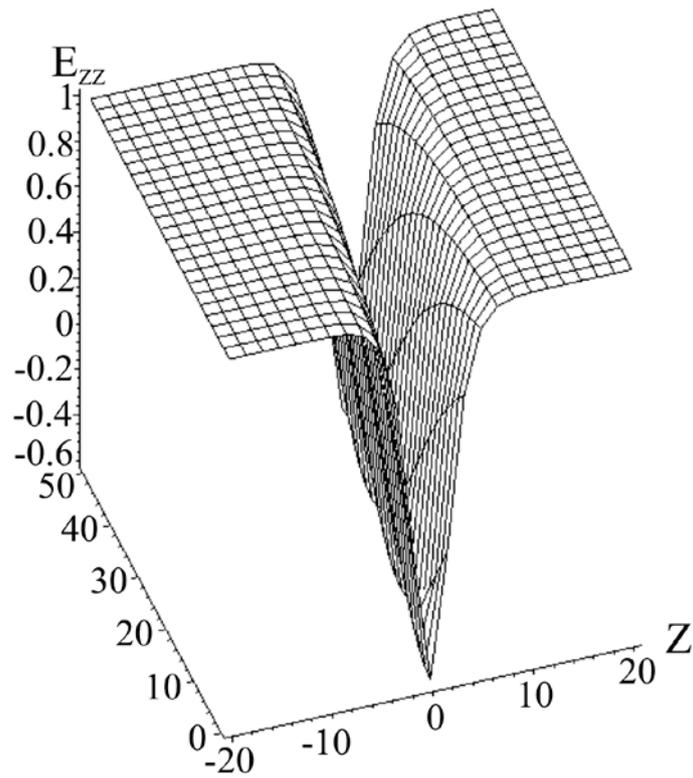
**Fig.7.** Non-linear deformation wave $E_{zz}$ along z. $\alpha=0.1$, $\beta=0.2$.

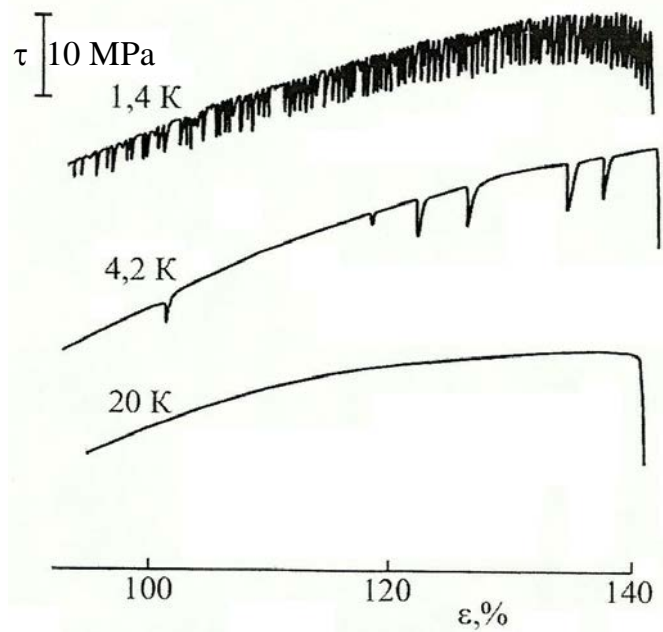
**Fig.8.** Sections of the tensile curve from 99.5 % -purity aluminum single crystal specimens at different temperatures [11]. The axis of tension is close to [111].

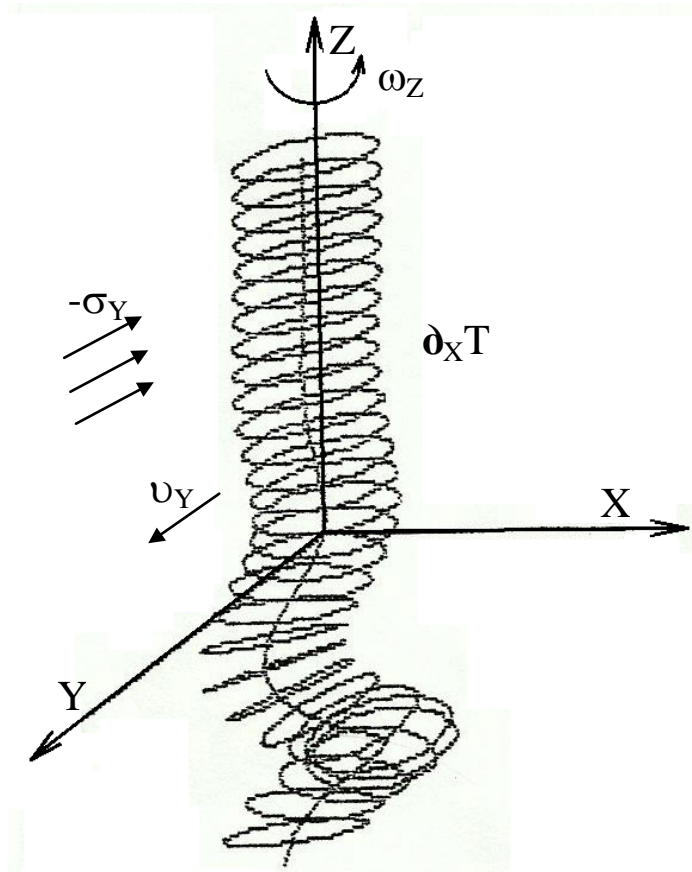

**Fig.9.** Helical wave of shape variation in the region of a whirlwind.

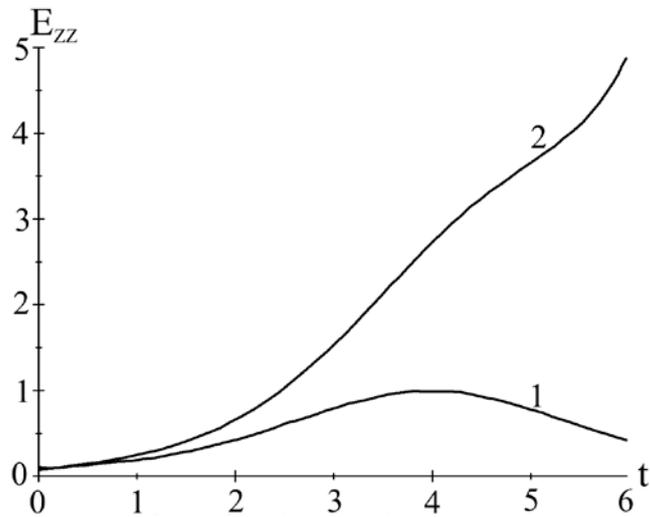

**Fig. 10.** Dependence of superplastic deformation (creep) on time. $L=2$, $\upsilon=0.5\alpha$.